\documentclass[onecolumn,aip, apl,preprint, floatfix]{revtex4-1}

\usepackage[dvips]{graphicx}
\usepackage{float}
\usepackage{siunitx}
\usepackage{color}
\usepackage{amsmath}
\usepackage{amssymb}
\usepackage[utf8x]{inputenc}
\usepackage{float}
\usepackage{blindtext}
\usepackage{xcolor}
\usepackage[version=3]{mhchem}

\definecolor{orange}{rgb}{1,0.5,0}

\newcommand{\uel}{\si{\micro}-EL}
\newcommand{\nm}{\nano\meter}

\draft 

\begin{document}


\title{Single-photon light emitting diodes based on pre-selected quantum dots using a deterministic lithography technique} 




\author{Marc Sartison}\email{m.sartison@ihfg.uni-stuttgart.de}\noaffiliation

\author{Simon Seyfferle}\noaffiliation

\author{Sascha Kolatschek}\noaffiliation

\author{Stefan Hepp}\noaffiliation

\author{Michael Jetter}\noaffiliation

\author{Peter Michler}\noaffiliation

\author{Simone Luca Portalupi}\noaffiliation

\affiliation{Institut f\"ur Halbleiteroptik und Funktionelle Grenzfl\"achen, Center for Integrated Quantum Science and Technology (IQ$^{ST}$) and SCoPE, University of Stuttgart, Allmandring 3, 70569 Stuttgart, Germany }


\date{\today}

\begin{abstract}
In the present study, we developed a fabrication process of an electrically driven single-photon LED based on InP QDs emitting in the red spectral range, the wavelength of interest coinciding with the high efficiency window of \ce{Si} APDs. 
A deterministic lithography technique allowed for the pre-selection of a suitable QD, here exclusively operated under electrical carrier injection.
The final device was characterized under micro-electroluminescence in direct current, as well as in pulsed excitation mode.
In particular, under pulsed excitation of one device, single-photon emission of a spectral line, identified as an exciton, has been observed with $g^{(2)}_\mathrm{raw}(0)=0.42\pm0.02$, where the non-zero $g^{(2)}$-value is mainly caused by background contribution in the spectrum and re-excitation processes due to the electrical pulse length.
The obtained results constitute an important step forward in the fabrication of electrically driven single-photon sources, where deterministic lithography techniques can be used to sensibly improve the device performances.
In principle, the developed process can be extended to any desired emitter wavelength above \SI{600}{\nm} up to the telecom bands.


\end{abstract}


\maketitle 




In the development of photonic-based quantum technology, several systems have been under discussion as sources of single and entangled photons.
Among several possibilities, semiconductor quantum dots (QDs) showed the capability of generating very pure single and indistinguishable photons\cite{Ding2016,Somaschi2016} as well as highly entangled photon pairs\cite{Huber2018}.
Despite state-of-the-art performances have
been reached under optical pumping, the possibility of implementing an
efficient electrical pumping scheme will be of key importance for
device scalability. An electrically-driven device would have an overall
smaller footprint and its performances would not be affected by
time dependent instabilities of the pumping conditions, as it may happen with a 
spatial or spectral drift of the excitation laser.
Electrically driven single-photon devices have been experimentally demonstrated, covering the spectrum from the blue up to the telecom regimes\cite{Yuan2002,Deshpande2013,Reischle2008,Kessler2012,Heindel2012,Ward2007,Muller2018}.
Additionally, pulse generators can be used for electrical carrier injection reaching on-demand operation\cite{Reischle2010,Zhang2015,Heindel2012} with high excitation repetition rates up to \SI{2}{\giga\hertz}\cite{Hargart2013}. 
Thus, high flexibility in terms of pumping conditions can be achieved, tailoring the emission properties to match the application needs.
This makes them a very versatile choice for complex experiments in the field of quantum information technology \cite{Knill2001,OBrien2009,Waks2002}
and metrology applications, which will largely benefit from a
light source which provides a stable and reliable photon flux.
Furthermore, since QDs are embedded in a semiconductor matrix, well established semiconductor processing techniques can be applied for their implementation into cavities\cite{Unsleber2015,Somaschi2016,Ates2012,Pelton2002,Gerard1998}
and various photonic structures\cite{Reimer2012,Claudon2010,Munsch2013,Trojak2017,Gschrey2015,Sartison2017,Sartison2018}  to enhance their performances.
For this purpose, deterministic fabrication techniques were developed
largely improving the fabrication yield due to the
capability to localize an emitter and tailor the
lithographically defined mask according to its emission properties\cite{Sartison2017,Sartison2018,Somaschi2016,Ates2012,Trojak2017,Gschrey2015,Unsleber2015}.
In this work, we based our device fabrication on in-situ photolithography aligned on an electrically pre-selected emitter. 
The use of electrical pumping, rather than the commonly utilized optical excitation, ensures that the emitters will have the desired performances after the device processing. 
Additionally, since optical excitation is not required, only two channels (collection and lithography) are necessary, in contrast to usual approaches \cite{Sartison2017,Dousse2008}.

Single-photon light-emitting diodes (SPLEDs) based on InP QDs emitting in the red
spectral range\cite{Kessler2012,Reischle2010} have shown their potential in realizing QKD experiments\cite{Heindel2012}.
Despite that, former demonstrations were limited by the non-deterministic device fabrication which turned out in SPLEDs including several QDs.
Only lowering the applied voltage allowed for the observation of emission from a single dot which is randomly placed into the large SPLED aperture.
Our newly developed method, enables us now to integrate electrically pre-selected QDs into similar optical devices while having a precise knowledge of the position\cite{Sartison2017} and the emission properties prior to the fabrication process. 
Furthermore, the device size could be reduced and a smaller aperture diameter resulted in a more concentrated current confinement which enables a more efficient pumping of the pre-selected QD.
As a proof of principle, we verified the non-classical emission properties of one pre-selected integrated QD under triggered electrical excitation.\\
The processed sample was grown via metal-organic vapor-phase epitaxy (MOVPE). 
It consists of a bottom n-doped distributed Bragg reflector (DBR) with 45 mirror pairs followed by the InP QD layer.
A high Al-containing oxidation layer was deposited before the structure was completed with a p-doped DBR containing 6 mirror pairs.
The oxidation layer is used during the device fabrication to ensure current confinement and selective excitation of individual QDs.
A detailed report on the sample structure can be found in Schulz et. al\cite{Schulz2011}. 

\begin{figure}[H]
	\centering
	\includegraphics{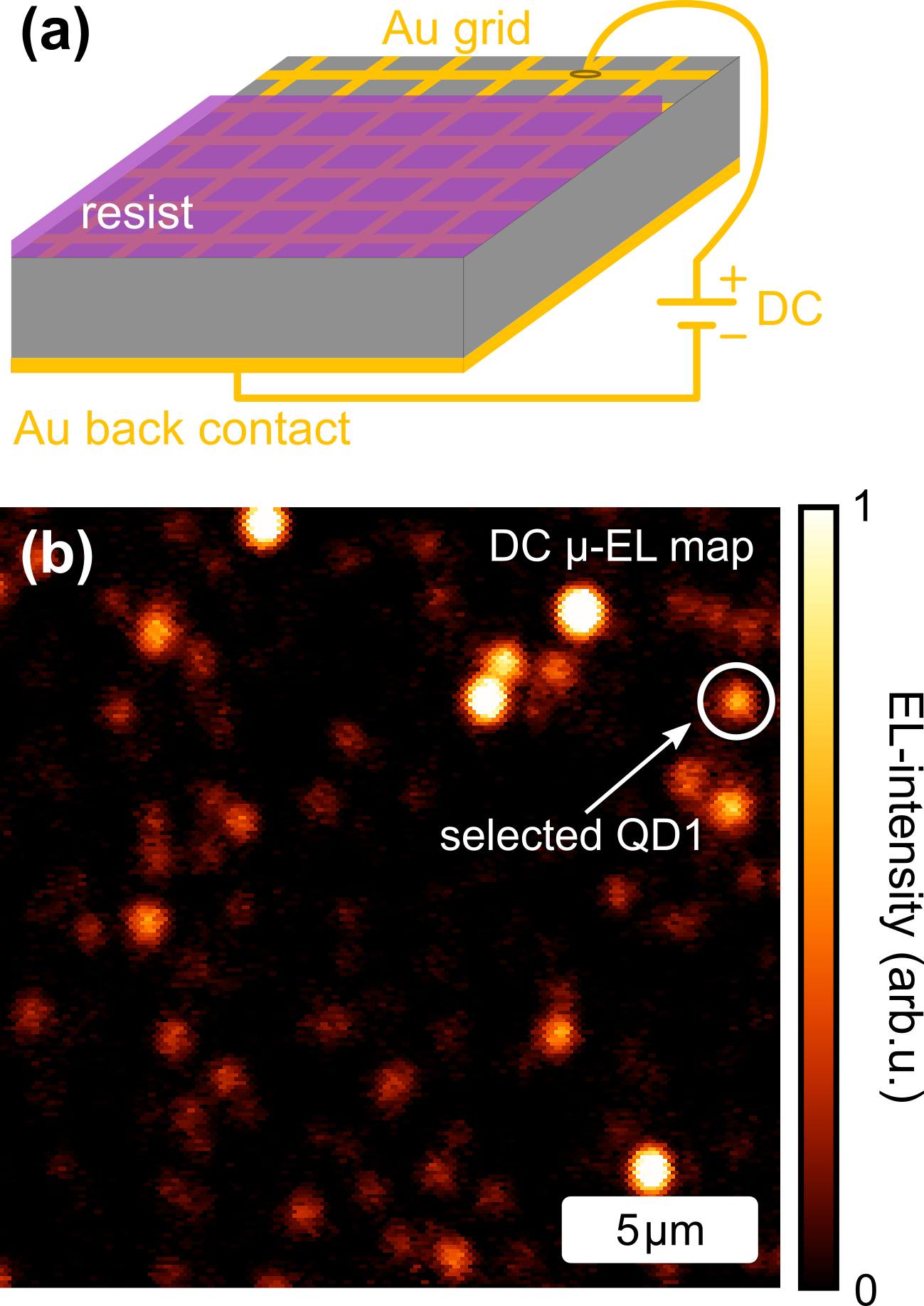}
	\caption{\textbf{(a)} Schematic of the contacted sample after being mounted in the cryostat. The resist is partially exposed and developed to create an opening for the depicted wire bond. \textbf{(b)} \uel~map under DC excitation. Since the detection is fiber coupled, the fiber serves as a pinhole for spatial selectivity. Single isolated emission spots are visible. The QD integrated in device\,$\#1$ is marked via the white circle.}
	\label{fig:fig1uel}
\end{figure}
\begin{figure}
	\centering
	\includegraphics{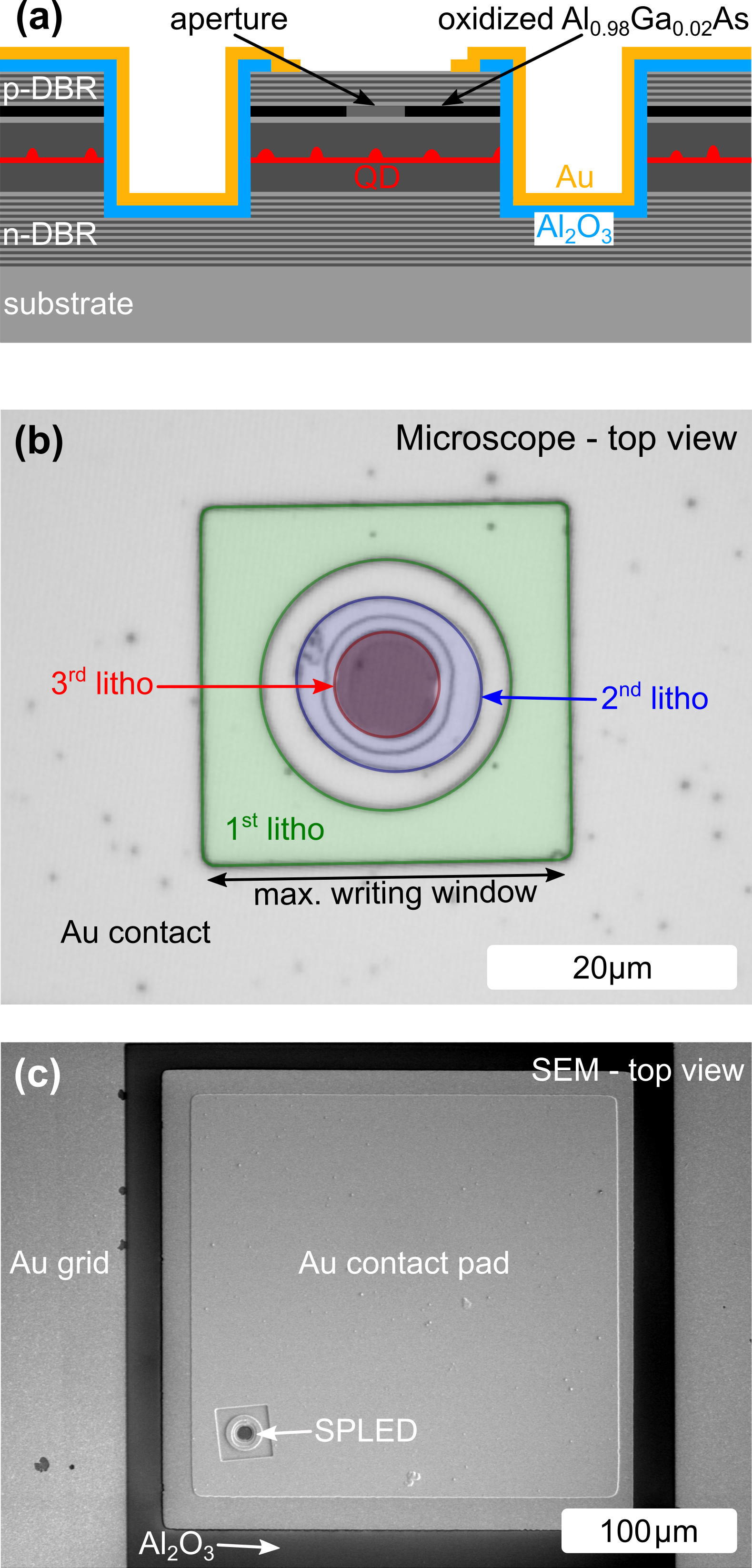}
	\caption{\textbf{(a)} Schematic of the fabricated SPLED. \textbf{(b)} Top view microscope picture of the final device. \textbf{(c)} Top view SEM picture of the device showing the SPLED, the \ce{Au} contact pad and the \ce{Al2O3} insulation layer underneath. Outside of the device, the initial \ce{Au} grid is also visible.}
	\label{fig:fig2device}
\end{figure}

As a first process step, a grid was lithographically defined via standard UV photolithography, followed by electron-beam evaporation of a \ce{Cr} and a \ce{Au} layer. 
The top of the sample then displays a grid similar to the one illustrated in Fig.\,\ref{fig:fig1uel}(a), while the bottom is fully covered by these metallic layers to constitute the bottom contact.
These contacts enable the observation of electroluminescence during the low-temperature deterministic lithography step.
For this purpose, the sample was again spin-coated with a positive resist, partially exposed and developed to get access to the contact grid on top.
After being mounted on an electrical sample holder, the partially exposed grid was contacted to the positive electrode via wire bonding (see sketched wire in Fig.\,\ref{fig:fig1uel}(a)). 
For QD pre-selection, a commercially available low-temperature photolithography setup was used as already described in former works\cite{Dousse2008,Sartison2017,Sartison2018}. 
However, this time the QD excitation was provided by supplying a direct current (DC) source instead of optical excitation.
This provides two advantages: on the one hand, it does not require the excitation laser to be carefully aligned with collection and lithography channels; on the other hand, it ensures that the pre-selected QD will also show emission in EL after the device processing.
When cooled down to \SI{4}{\kelvin},  \uel~maps were acquired under \SI{2.5}{\volt} DC excitation. Bright and isolated QD emission could be observed, despite a modest rise in sample temperature to \SI{5.6}{\kelvin} (see Fig.\,\ref{fig:fig1uel}(b)). 
This is due to the large current injection area but only has a modest effect on the dot emission properties\cite{Reischle2008}. 

\begin{figure}[h]
	\centering
	\includegraphics{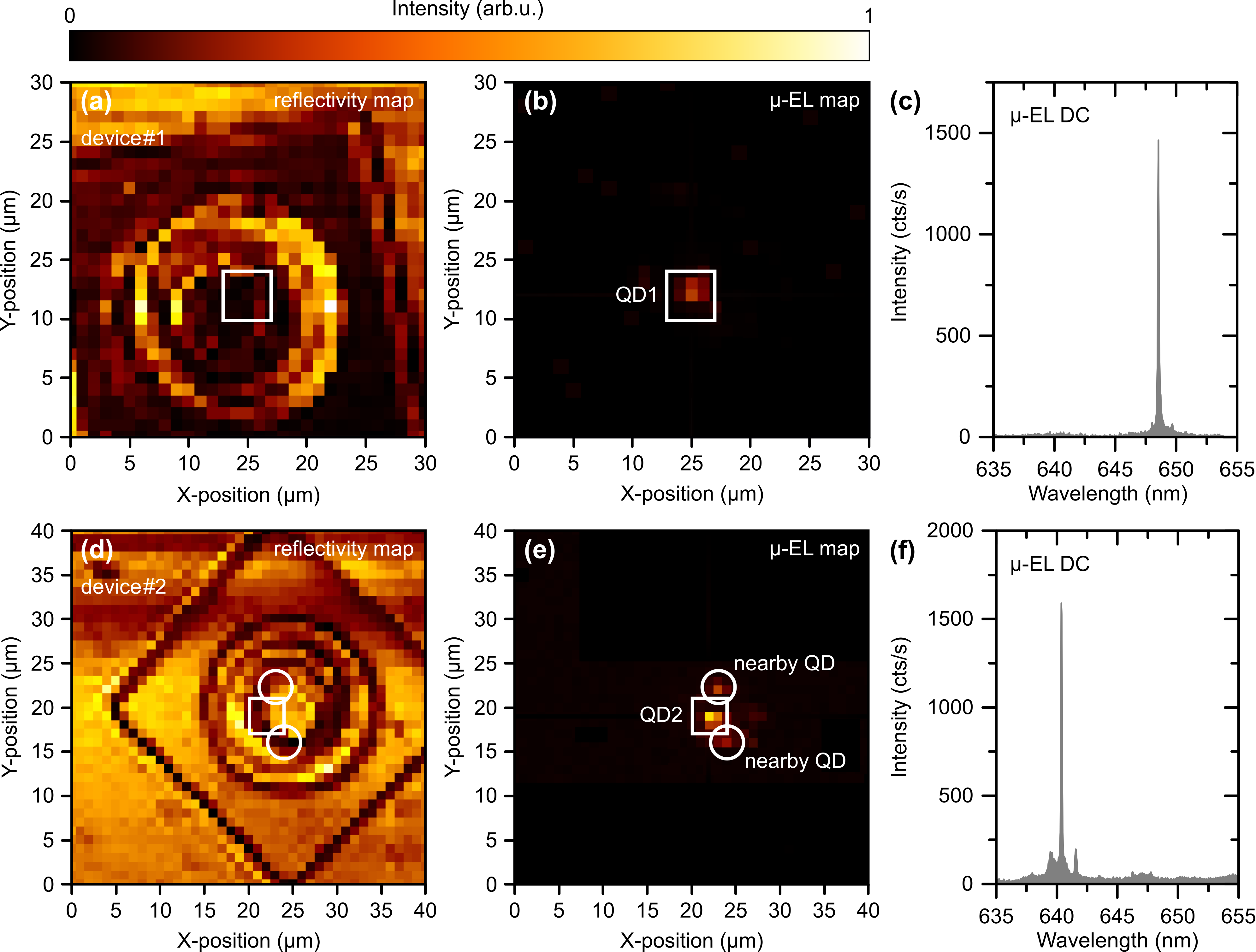}
	\caption{\textbf{(a)} Reflectivity scan of device\,$\#1$. \textbf{(b)} \uel~map of device\,$\#1$. Emission of only one QD is visible, namely QD1. \textbf{(c)} QD1 corresponding \uel~spectrum.
		\textbf{(d)} Reflectivity scan of device\,$\#2$. \textbf{(e)} \uel~map of device\,$\#2$. Emission of multiple QDs is visible, namely the pre-selected QD2 and two further QDs which are nearby. \textbf{(f)} QD2 corresponding \uel~spectrum. \\
		Pre-selected QDs are marked with a square which yield also the brightest emission intensity. QDs nearby are marked with a circle. A fiber in the detection path was used as a pinhole for enhanced spatial resolution.}
	\label{fig:fig3mapsundspektrumvergleichv2}
\end{figure}

As soon as a spatially and spectrally isolated QD is identified, the device is processed, deterministically aligned on the spatial position of the QD, as described in the following (a detailed sketch of the fabrication steps can be found in Fig.\,S1 in the supplementary material). 
Fig.\,\ref{fig:fig2device}(a) illustrates the final device in cross section.
In order to realize such a structure, a writing window of $\SI{30}{\um}\times\SI{30}{\um}$ was exposed at low temperature around the QD except a circle with a diameter of approximately \SI{20}{\um} (marked as $1^\mathrm{st}$ litho in Fig.\,\ref{fig:fig2device}(b)). The unexposed resist which remains after the development served as an etching mask which was transferred into the sample structure via inductively-coupled reactive-ion etching using \ce{SiCl4}:\ce{Ar}.
The etching should not be stopped before reaching the bottom DBR. 
This gives lateral access to the high aluminum containing layer, providing the possibility of later wet-chemical aperture oxidation.
At this point, the sample was covered again with resist, this time with an image reversible one, and placed in the in-situ lithography setup. 
Since the position of the QD is already marked by the device itself, i.e. the emitter is located in the center of the etched structure, the second lithography could be carried out at room temperature.
A red laser at \SI{658}{\nm}, aligned to be collinear with the lithography channel, was scanned over the device and the reflectivity was mapped via a photodiode using the back reflected laser signal (see Fig.\,S2 in the supplementary material). 
Based on this map, the position and size of the circle to be exposed was determined. 
Here, the diameter of the second lithography (Fig.\,\ref{fig:fig2device}(b) $2^\mathrm{nd}$ litho) was set to approximately \SI{13}{\um}.
This size has to be smaller than the circle of the first lithography to ensure electrical insulation provided by the subsequently deposited \ce{Al2O3} layer anywhere except in the center of the device (see blue layer in Fig.\,\ref{fig:fig2device}(a)). 
After the image reversal bake, flood exposure and development of the resist, \ce{Al2O3} was deposited and a lift-off step was performed. 
This allows for the realization of an insulating layer all over the sample, except in the device central area, where the QD is sitting. In the following step,
wet-chemical oxidation defined an aperture in the high aluminum containing layer of approximately \SI{8}{\um} in order to further limit the area of current flow (see Fig.\,\ref{fig:fig2device}(a)).
QDs outside this aperture are not reached by the injected carriers.
Optimally, when the QD density is low enough, it is possible to obtain emission of only one QD when the device is operated.
The previously described lithographic step was repeated, but this time the exposed circle ($3^\mathrm{rd}$ litho Fig.\,\ref{fig:fig2device}(b)) had to be smaller than the opening in the \ce{Al2O3} (exposure diameter chosen to be around \SI{8}{\um}).
A grid was also exposed in standard UV lithography around the structures which defines the contact area for addressing each device individually via wire bonding.
It followed the resist development, the contact deposition using \ce{Cr} and \ce{Au} layers and a lift-off step.
Figure\,\ref{fig:fig2device}(c) shows the device including the large gold contact pad, which is needed for wire bonding.
After device fabrication, the sample was placed on a custom sample holder, suitable for DC and high-frequency pulsed electrical operation, and contacted via wire bonding. 
Thereafter, it was cooled down to \SI{4}{\kelvin} in a helium flow cryostat. 
The first characterization step was devoted to verifying the successful fabrication. For this scope, reflectivity and electroluminescence in DC were simultaneously recorded. 
A very weak laser signal from a He-Ne laser was added to the DC bias during the mapping, which does not modify the electrically-pumped QD emission.
This enabled to gather information about device topography (in reflectivity) and electrical emission (\uel) at the same time.
By applying a spectral filter in the evaluation of the spectra, either the topography (Fig.\,\ref{fig:fig3mapsundspektrumvergleichv2}(a) and (d)) or the QD emission (Fig.\,\ref{fig:fig3mapsundspektrumvergleichv2}(b) and (e)) could be extracted.
Each visible pixel corresponds to a recorded spectrum, as exemplary shown in Fig.\,\ref{fig:fig3mapsundspektrumvergleichv2}.
The respective QD position, within the collection spot size of around \SI{4}{\micro\meter} is marked in all maps, reflectivity and \uel.
Here, a fiber was used in the detection path serving as a pinhole for a higher spatial resolution.
For device\,$\#1$, only the pre-selected QD was emitting inside the aperture. 
The corresponding spectrum is shown in Fig.\,\ref{fig:fig3mapsundspektrumvergleichv2}(c). 
Device\,$\#2$ showed emission of two additional QDs together with the pre-selected one. However, the selected QD was the most central one as expected. 
In Fig.\,\ref{fig:fig3mapsundspektrumvergleichv2}(f), the spectrum of QD2, the pre-selected one, is shown. 
It is important to mention that the oxidation step induced a significant compressive strain on the QD centered underneath the oxide aperture which resulted in a blue-shift of  \SI{17.3}{\nm} and \SI{16.4}{\nm} for devices $\#1$ and $\#2$, respectively.
Thanks to this aperture, which forces the current flow around the QD, no detectable change of the device temperature was observed, differently from the operation during the first low-temperature lithography step.

\begin{figure}[h]
	\centering
	\includegraphics[scale=1]{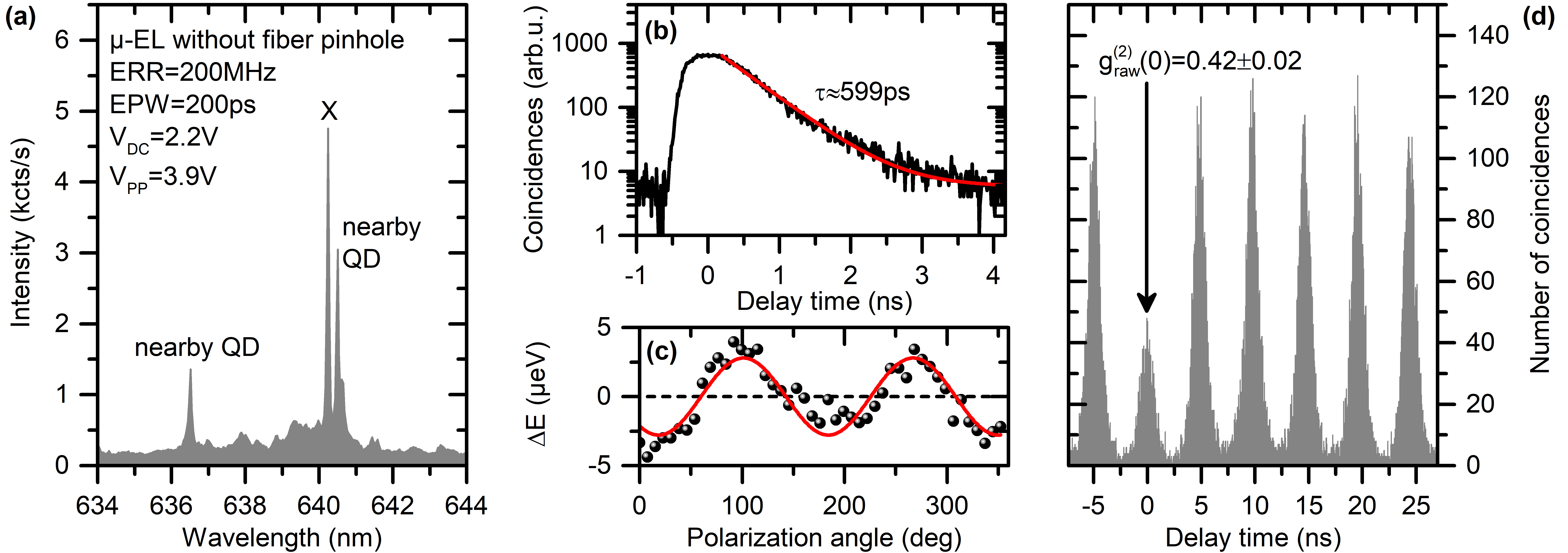}
	\caption{Optical characterization of the SPLED device\,$\#2$: \textbf{(a)} \uel~spectrum with excitation parameters $\mathrm{ERR}=\SI{200}{\mega\hertz}$, $\mathrm{EPW}= \SI{200}{\pico\second}$, $\mathrm{V}_\mathrm{DC}= \SI{2.2}{\volt}$, $V_{\text{pp}}=\SI{3.9}{\volt}$. 
	\textbf{(b)} Time-resolved measurement of the bright emission line marked as X in (a) showing a decay time of \SI{599}{\pico\second}.
	\textbf{(c)} Polarization dependent measurement of the X line in (a) resulting in a periodic energetic oscillation around the central line with an amplitude of \SI{2.79}{\micro\electronvolt}.
\textbf{(d)} Raw data second-order autocorrelation measurement performed on the emission line marked as X in (a) filtered within a spectral window of $\Delta \lambda =\SI{60}{\pico\meter}$. The suppression of the peak at zero delay time indicates single-photon emission.}
	\label{fig:g2}
\end{figure}

Pulsed electrical pumping of the device was accomplished by a pulse generator suitable for narrow excitation pulse widths (EPW) and high excitation repetition rates (ERR). Its output was coupled to an amplifier to increase the peak voltage ($\mathrm{V}_\mathrm{pp}$) up to a maximum of 5\,V. With the use of a bias tee, an additional constant DC bias voltage ($\mathrm{V}_\mathrm{DC}$) could be applied which was set slightly below the excitation threshold.\\
Pulsed excitation of device\,2 with an ERR of \SI{200}{\mega\hertz} and a pulse width of \SI{200}{\pico\second} resulted in the spectrum displayed in Fig.\,\ref{fig:g2}(a). 
In contrast to the map and spectrum in Fig.\,\ref{fig:fig3mapsundspektrumvergleichv2}, no spatial filter was used here, i.e. light was sent in free space to the spectrometer to maximize the count rate on the APDs.
As a consequence, the EL intensity of the nearby QDs is slightly increased with respect to the intensity of the pre-selected QD (compare Fig.\,\ref{fig:fig3mapsundspektrumvergleichv2}(f) and Fig.\,\ref{fig:g2}(a)).
The AC peak voltage was set to $\mathrm{V_{pp}}=\SI{3.9}{\volt}$ while the DC bias was kept at \SI{2.2}{\volt}.
For the upcoming detailed measurements, the brightest emission line in Fig.\,\ref{fig:g2} was chosen.
Time-correlated single-photon counting (TCSPC) revealed a decay time of $\tau=\SI{599}{\pico\second}$ (see Fig.\,\ref{fig:g2}(b)), a value typical for electrically driven InP QDs\cite{Reischle2010}. 
Polarization dependent measurements were performed on this line, which showed a  spectral oscillation with an amplitude of $\SI{2.79}{\micro\electronvolt}$ around the central peak position marked as the dashed line in Fig.\,\ref{fig:g2}(c)).
This oscillation is an indication of a fine structure splitting of the line.
Considering the decay time of \SI{599}{\pico\second}, the conclusion can be drawn that this line stems from an excitonic transition.
A second-order autocorrelation measurement on this emission line resulted in the histogram seen in Fig.\,\ref{fig:g2}(d).  Integration of the coincidences within a temporal window of \SI{2}{\nano\second} gave a raw value of $g^{(2)}_\mathrm{raw}(0)=0.42\pm0.02$
and a dark count corrected value of $g^{(2)}_\mathrm{dcc}(0)=0.41\pm0.03$.
The non-perfect suppression of the zero-delay peak can be accounted to residual background emission as well as possible re-excitation during the same pulse. 
A benchmark for the signal $S$ to background $B$ ratio $\rho =S/(S+B)$ could be obtained from fitting the spectrum to distinguish between the QD signal and the underlying background (see Kessler et al.\cite{Kessler2012}). 
Thus, the $g^{(2)}_\mathrm{dcc}(0)$ could be background corrected via the formula\cite{Brouri2000} $g^{(2)}_{\text{bgc}}(0)=(g^{(2)}_\mathrm{dcc}(0)-(1-\rho ^2))/\rho ^2$,
resulting in a value of $g^{(2)}_{\text{bgc}}(0)=0.23\pm 0.05$
which compares with previously obtained results\cite{Kessler2012} achieved with similar excitation pulse lengths.
Additionally, it has to be considered that the non-negligible tails in the Gaussian excitation pulse may induce re-excitation of the QD with consequent degradation of the $g^{(2)}(0)$. 
Therefore, future measurements have to be performed using pulses with an EPW much shorter than \SI{200}{\pico\second}  in order to avoid re-excitation.\\
In conclusion, we realized a QD-based single-photon LED, operating at the absolute efficiency maximum of Si APDs, via an optimized deterministic fabrication technique. 
QDs were here pre-selected in \uel~and the device was fabricated via state-of-the-art low temperature in-situ lithography and standard clean room fabrication techniques.
The successful deterministic integration of QDs could be verified by the acquisition of \uel~maps combined with reflectivity maps.
Furthermore, the device performances were characterized under pulsed electrical excitation by means of EL polarization dependence, time-resolved EL decay time and photon-autocorrelation measurements. 
A decay time of \SI{599}{\pico\second} under triggered electrical excitation compares well with already reported values for \ce{InP} QDs in literature.
Additionally, a $g^{(2)}_\mathrm{dcc}(0)$ value of $0.41\pm0.03$ could be recorded, only corrected for the detector induced dark count contribution, thus indicating dominant single-photon emission.
By accounting also for the spectral background contribution, a corrected $g^{(2)}_{\mathrm{bgc}}(0)$ value of $0.23\pm 0.05$ could be estimated.
The deterministic fabrication process of SPLEDs developed in this work can be applied to any electrically driven emitter from the red spectral range up to the telecom regimes. 
The proof of principle measurements presented here show the possibility of combining in-situ optical lithography with electrically driven non-classical light sources. It enables the route towards electrically driven cavity systems and high brightness devices.
Improving the performances of single QD devices via deterministic lithography techniques will be of key importance in a range of quantum technology implementations, from quantum key distribution to quantum enhanced metrology.
\vspace{\baselineskip}

The authors would like to thank the DFG for financial support via the project Mi500/27-1. 
S.K. and S.L.P. greatly acknowledge the Baden-W\"urttemberg Stiftung ``Post-Doc Elite Programm'' via the project ``Hybride Quantensysteme f\"ur Quantensensorik''. The research of the $\text{IQ}^{\text{ST}}$ was financially supported by the Ministry of Science, Research and Arts Baden-W\"urttemberg.


\bibliography{mybib.bib}
\bibliographystyle{apsrev4-1}

\end{document}



\title{Supplementary information: Single-photon light emitting diodes based on pre-selected quantum dots using a deterministic lithography technique.} 




\author{Marc Sartison}\email{m.sartison@ihfg.uni-stuttgart.de}\noaffiliation

\author{Simon Seyfferle}\noaffiliation

\author{Sascha Kolatschek}\noaffiliation

\author{Stefan Hepp}\noaffiliation

\author{Michael Jetter}\noaffiliation

\author{Peter Michler}\noaffiliation

\author{Simone Luca Portalupi}\noaffiliation

\affiliation{Institut f\"ur Halbleiteroptik und Funktionelle Grenzfl\"achen, Center for Integrated Quantum Science and Technology (IQ$^{ST}$) and SCoPE, University of Stuttgart, Allmandring 3, 70569 Stuttgart, Germany }



\pacs{}

\maketitle
\begin{figure}[H]
	\centering
	\includegraphics[]{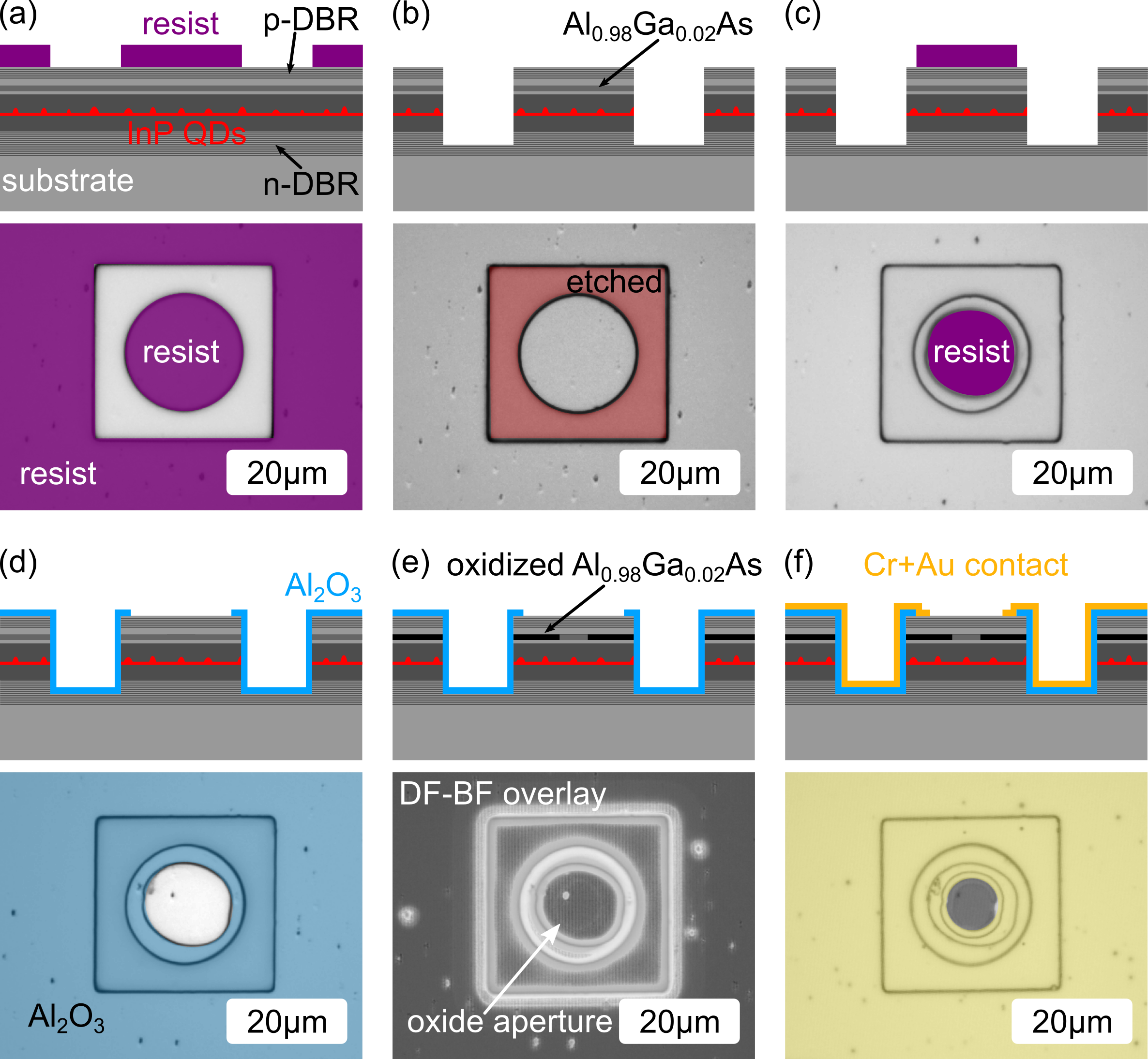}
	\caption{Cross-section illustrations and microscope top view pictures of various single-photon LED fabrication steps. \textbf{(a)} After deterministic low temperature in-situ lithography.
	\textbf{(b)} After ICP-RIE etching and resist removal. 
	\textbf{(c)} After image reversal spin-coating, aligned in-situ room temperature deterministic lithography and development.
	\textbf{(d)} After \ce{Al2O3} deposition and lift-off.
	\textbf{(e)} Bright field and dark field overlay after wet chemical oxidation. The inner shallow white ring shows the oxide aperture.
	\textbf{(f)} Final device after electrical contact deposition.}
	\label{fig:prozesspng600dpi}
\end{figure}

\begin{figure}[H]
	\centering
	\includegraphics{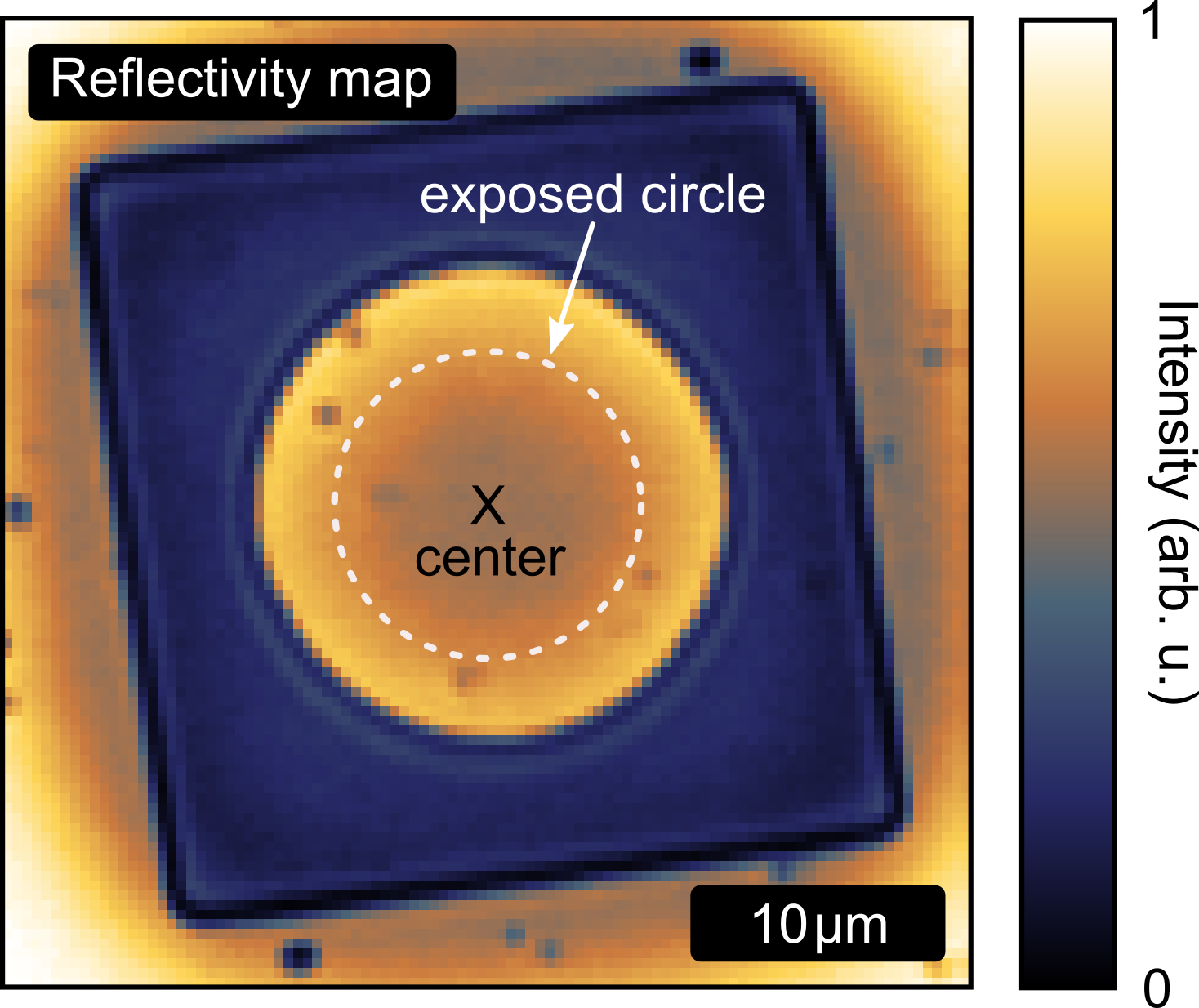}
	\caption{Reflectivity map of the etched device from \fig\ref{fig:prozesspng600dpi}(b) already spin-coated with resist. The center of alignment of the following adapted lithography is marked by the X while the nominal shape to be exposed is illustrated by the dashed circle.}
	\label{fig:rcledalignmentinsitu}
\end{figure}

 
In here, we give a more detailed insight into the used fabrication steps.
The sample preparation up to the deterministic lithography is applied as described in the manuscript.
Figure\,\ref{fig:prozesspng600dpi}(a)-(f) show the process steps after the in-situ exposure in more detail. For each stage, a cross-sectional sketch corresponds to the colored top view microscope pictures, respectively. 
When a QD with desired characteristics is located, a window of $\SI{30}{\um} \times \SI{30}{\um}$ is exposed except a circle with a diameter of around \SI{20}{\um}. After development, the structure takes the shape shown in the microscope picture in \fig\ref{fig:prozesspng600dpi}(a). It follows an ICP-RIE etching step with gases composed of \ce{SiCl4} and \ce{Ar}. 
The etching is stopped after entering the n-DBR layers. This etching depth can be in-situ monitored on a separate similar sample by a built-in interferometric measurement setup of the commercially-available Oxford Plasma Lab 100.
After the etching, the high Al-containing layer remains uncovered on the flank around central area (see \fig\ref{fig:prozesspng600dpi}(b)). Here, the etched region is marked in the microscope picture in red.
It follows another in-situ laser lithography step at room temperature. This time, an image reversal photoresist is used. 
The lithography mask is aligned by acquisition of a reflectivity map of the etched structure and the origin and the size of the circle to be written are adapted accordingly. An example of such a reflectivity map can be seen in \fig\ref{fig:rcledalignmentinsitu}.
The middle of the central area is marked by the X and a circle with a diameter of \SI{13}{\um} is depicted in white dashed.
Exposure of this region is followed by an image reversal bake and a flood exposure to invert the resists behavior under development. When developing, only the exposed part remains, as shown in \fig\ref{fig:prozesspng600dpi}(c). 
In order to insulate the sample surface, except an area centered on the device, \ce{Al2O3} is deposited via electron-beam evaporation. After a lift-off step, the sample looks like in \fig\ref{fig:prozesspng600dpi}(d). The deposited \ce{Al2O3} layer is here depicted in blue.
Even though sidewalls are now covered with \ce{Al2O3}, the \ce{Al_{0.98}Ga_{0.02}As} layer can be oxidized to define an oxide aperture for limiting the current flow. For this purpose, the sample is placed inside a custom wet-chemical oxidation oven at \SI{360}{\degreeCelsius}. 
The oxidation status, thus the aperture size, can be monitored in-situ with a built-in microscope. 
Figure\,\ref{fig:prozesspng600dpi}(e) shows a dark field (DK) and bright field (BF) overlay image of the structure to show information about the sample geometry and the oxide aperture at the same time.
The white arrow points towards the edge of the oxidized layer, leaving an aperture size of approximately \SI{8}{\um} for an oxidation time of \SI{3000}{\second}. 
For obtaining the final device, depicted in \fig\ref{fig:prozesspng600dpi}(f), step (c) and (d) are repeated, but with an even smaller exposed circle (diameter approximately \SI{8}{\um}) and a deposition of \ce{Cr} and \ce{Au} instead of \ce{Al2O3}, which completes the structure.

